\documentclass[a4paper,11pt]{article}
\pdfoutput=1 

\usepackage{jcappub} 
\usepackage{subfigure}
\usepackage{graphicx}
\usepackage[T1]{fontenc} 

\title{\boldmath Gravitational radiation from binary systems in $f(R)$ gravity: A semi-classical approach}

\author[a,c]{Ashish Narang,}
\author[c]{Subhendra Mohanty,}
\author[b,1]{Soumya Jana,\note{Corresponding Author. Email: soumyajana.physics@gmail.com}}

\affiliation[a]{Institute of Physics, Bhubaneswar, 751005, India}
\affiliation[b]{Department of Physics, Sitananda College,  Nandigram, 721631, India}
\affiliation[c]{Theoretical Physics Division, Physical Research Laboratory, Ahmedabad, 380009 India}

\emailAdd{anarangg@gmail.com}
\emailAdd{mohanty@prl.res.in}
\emailAdd{soumyajana.physics@gmail.com}

\abstract{ The rate of energy loss and orbital period decay of quasi- stable compact binary systems are derived in $f(R)$ theory of gravity using the method of a single vertex graviton emission process from a classical source. After linearising the $f(R)$ action written in an equivalent scalar-tensor format in the Einstein frame, we identify the appropriate interaction terms between the massless spin-2 tensor mode, massive scalar mode, and the energy momentum tensor. The definition of the scalar field is related to the $f(R)$ models. Then using the interaction vertex we compute the rate of energy loss due to spin-2 quadrupole radiation, which comes out to be the same as the Peter-Mathews formula with a multiplication factor, and also the energy loss due to the scalar dipole radiation. The total energy loss is the sum of these two contributions. Our derivation is most general as it is applicable for both  arbitrary eccentricity of the binary orbits  and arbitrary mass of the scalar field. Using the derived theoretical formula for the period decay of the binary systems, we compare the predictions of $f(R)$ gravity and general relativity for the observations of four binary systems, i.e. Hulse-Taylor Binary, PSR J1141-6545, PSR J1738+0333, and PSR J0348+0432. Thus we put bound on three well-known $f(R)$ dark energy models, namely the Hu-Sawicki, the Starobinsky, and the Tsujikawa model. We get the best constraint on $f'(R_0)-1$ (where $R_0$ is the scalar curvature of the Universe at the present epoch) from the Tsujikawa model, i.e $\vert f'(R_0)-1\vert < 2.09\times 10^{-4}$. This bound is stronger than those from most of the astrophysical observations and even some cosmological observations.}


\begin{document}
\maketitle
\flushbottom

\section{Introduction}

General relativity (GR), since its inception in 1916, is the most successful classical theory of gravity which has passed all experimental tests \cite{Weinberg:1972kfs} and more so with the recent direct detection of gravitational waves by the LIGO-Virgo collaboration \cite{PhysRevLett.116.061102} and the observation of the black hole images by the Event Horizon Telescope \cite{Akiyama_2019}. 
However, some long-standing puzzles in GR are still not solved and understood with full consistency. One such puzzle is understanding the observed accelerated expansion of the Universe which motivate many researchers to pursue modified gravity theories in the classical framework with deviation from GR in the infrared energy scales. The simplest modification in this direction is the $f(R)$ theory of gravity which is a generalization of the Einstein-Hilbert action by replacing the Ricci scalar ($R$) with a function $f(R)$ (see \cite{DeFelice:2010aj,NOJIRI201159,NOJIRI20171} and the references therein for a review). In some $f(R)$ models, the cosmological constant and  the dark energy, i.e., a new exotic form of matter are not necessary for the late time acceleration of the Universe. Initial versions of such models~\cite{Capozziello:2002rd,Capozziello:2003gx,Capozziello:2003tk,PhysRevD.70.043528} usually suffer from various instability problems~\cite{DOLGOV20031,PhysRevD.74.104017,PhysRevD.75.127502,PhysRevD.75.044004} and also do not satisfy the local gravity constraints~\cite{CHIBA20031,PhysRevLett.95.261102,PhysRevD.72.083505,PhysRevD.74.121501}. Later, several class of $f(R)$-dark energy models were proposed \cite{PhysRevD.68.123512,PhysRevD.74.086005,Hu:2007nk,Starobinsky:2007hu,PhysRevD.77.023507,PhysRevD.77.026007,PhysRevD.77.046009} which are stable and do satisfy cosmological and solar-system constraints under certain limits on the parameter space. In general, $f(R)$ gravity carries a massive scalar degree of freedom apart from the usual massless spin-2 tensor modes \cite{CAPOZZIELLO2008255,PhysRevD.95.104034}. Dynamically, $f(R)$ gravity is equivalent to Einstein gravity minimally coupled to a scalar field  in the Einstein frame, where the scalar field is associated with a nontrivial potential that depends upon the form of the $f(R)$ and couples to matter through the trace of the energy-momentum tensor. In the non-relativistic limit, the scalar field sources a (finite-range) fifth force which is added to the usual Newtonian force. The role of this extra scalar field in gravitational radiation and weak-field metric for simple sources was studied in \cite{berry} using the linearized form of $f(R)$ gravity. 
In some $f(R)$ theories, the fifth force can be screened only at the galactic or Solar-System scales through the chameleon mechanism \cite{Hu:2007nk,PhysRevD.69.044026,Burrage:2017qrf}. This mechanism facilitates the above mentioned viable models to conform the local gravity constraints as well as the modified dynamics at the large scale. Recently, in Refs.~\cite{smg_gw,PhysRevD.98.083023}, the authors have discussed how such screening mechanisms in scalar-tensor theories affect the gravitational radiation from compact binary systems.

It is interesting to ask what will happen if mass of the scalar field vanishes or more precisely, is there any vDVZ discontinuity in the $f(R)$ theory? The quest is originally linked to the Fierz-Pauli (FP) theory of massive gravity \cite{doi:10.1098/rspa.1939.0140}. Generically, in a massive field theory a particle exchange gives rise to a $(1/r)e^{-m_g r}$ Yukawa potential which goes to the $1/r$ potential in the $m_g\rightarrow 0$ limit. The FP theory of massive graviton has the peculiarity that in the zero graviton-mass limit the Lagrangian goes smoothly to Einstein-Hilbert (EH) linearized gravity theory, while the graviton propagator has additional contributions from the scalar modes of the metric which do not decouple in the zero graviton mass limit. As a result, the Newtonian potential in the zero-mass limit of FP theory is a factor $(4/3)$ larger than the prediction from the EH theory (which of course agrees with the Newtonian potential). The $f(R)$ theory also shares the same story. This peculiarity of the FP theory where the action goes to the EH theory in the zero mass limit but the graviton propagator does not, was first pointed out by van Dam and Veltman \cite{vanDam:1970vg} and independently by Zakharov \cite{Zakharov:1970cc} and this feature which arises in most massive gravity theories \cite{Hinterbichler:2011tt,deRham:2014zqa, Mitsou:2015yfa,Joyce:2014kja} is called the van Dam-Veltman-Zakharov (vDVZ) discontinuity (although, in the nonlinear FP theory, a proper decoupling limit will display the vDVZ discontinuity already in the action).
However, $f(R)$ does not turn into EH action in the zero mass limit of the scalar field unlike FP theory. Thus the anomaly of the zero mass limit in $f(R)$ theory is different than the vDVZ discontinuity of the other massive gravity theories. This anomaly can be avoided in some class of $f(R)$ theories which admit the  Chameleon screening mechanism \cite{Burrage:2017qrf,Hu:2007nk,PhysRevD.104.084017}.    

Previously, gravitational radiation from compact binary systems  in $f(R)$ theories has been studied \cite{smg_gw,PhysRevD.98.083023,PhysRevD.99.044056}. Compact binary systems are excellent laboratories to test theories of gravity in the strong field regime. The first indirect evidence of gravitational wave (GW) radiation was obtained from precision observations of the Hulse-Taylor binary system \cite{Hulse:1974eb,Taylor:1982zz,Weisberg:1984zz}. The orbital period loss of the compact binary system confirms Einstein's GR  \cite{Peters:1963ux} to $\sim 0.1 \% $ accuracy \cite{Weisberg:2016jye}. Following the Hulse-Taylor binary there have been other precision observation from compact binary systems \cite{Kramer:2006nb,Antoniadis:2013pzd,Freire:2012mg}. 

The objective of the present paper is to derive the rate of energy loss of binary systems (in quasi-stable orbits) due to gravitational radiation in observationally viable $f(R)$ theories using a one-graviton vertex process and compare the predictions of GR and the predictions of $f(R)$ theories. Earlier in \cite{PhysRevD.54.1474}, authors derived the rate of period decay of the binary pulsar systems by using entirely classical method. In~\cite{LIU2018286} the period decay rate was computed classically and was used to constrain various $f(R)$ models using observed period decay of the binary pulsar systems. However, the authors did not consider eccentricity in their calculation. 
GR in the weak field limit can be treated as a quantum field theory of spin-2 fields in the Minkowski space \cite{Feynman:1996kb, Weinberg:1964ew,Veltman:1975vx, Donoghue:2017pgk}.  Any classical gravity interaction like Newtonian potential between massive bodies or bending of light by a massive body can be described by a tree level graviton exchange. The result of the tree level exchange should match the weak field classical GR results. The derivation of gravitational radiation from binary stars as a single vertex Feynman diagram of massless graviton emission from a classical source has been performed in \cite{Mohanty:1994yi,Mohanty:2020pfa} and the results match with the result of Peter and Mathews~\cite{Peters:1963ux} who used the quadrupole formula of classical GR. This method was also used in computation of other radiations such as the vector gauge boson radiation \cite{KumarPoddar:2019ceq} and the massive graviton radiation \cite{Poddar2022}. In section~\ref{sec:lingra}, the linearized $f(R)$ action is derived in the scalar-tensor form. Using the linearized action, the rate of energy loss of a binary system due to gravitational radiation is obtained in section~\ref{sec:rad}. Using the theoretical results,  observational constraints on three well known $f(R)$ dark energy models, i.e., Hu-Sawicki, Starobinsky, and Tsujikawa models are also obtained for four NS-NS/WD binary systems in section~\ref{sec:constr}. In section~\ref{sec:conc}, we present our results. 

Throughout the paper we have used the natural system of units: $\hbar=c=1$, and $8\pi G=1/{M^2_{pl}}$ where Planck mass $M_{pl}=2.435\times 10^{18}$ GeV.

\label{sec:intro}

\section{Linearized f(R) action}\label{sec:lingra}
The $f(R)$ gravitational action is given by,
\begin{equation}
    S=\int \sqrt{-g} \mathrm{d^4 x}\left [-\frac{1}{16 \pi G}f(R)+\mathcal{L}_{M}\right ]
    \label{f(R) action in jordan frame}
\end{equation}
where $\mathcal{L}_M$ is the matter part of the action. Expanding the function $f(R)$ in the Taylor series around $R=0$, we get
\begin{equation}
    f(R)=f(0)+f^{\prime}(0)R+\frac{f^{\prime \prime}(0)}{2!}R^2+\mathcal{O}(R^3)+...
\end{equation}
Assuming $f(0)=0$, $f^{\prime}(0)=1$ and $\frac{1}{2} f^{\prime \prime}(0)=\alpha$, the series becomes
\begin{equation}
    f(R)=R+\alpha R^2+\mathcal{O}(R^3)+...
\end{equation}
Next we expand the gravitational action with respect to $g_{\mu \nu}=\eta_{\mu \nu}+\kappa h_{\mu \nu}$ and $\kappa=\sqrt{32 \pi G}$, where $ h_{\mu\nu}$ is the gravitation wave propagating in the flat the background. 
The Ricci tensor and the Ricci scalar are expanded upto first order in $h_{\mu\nu}$ as
\begin{eqnarray}
    R_{\mu \nu}&=&\frac{\kappa}{2}\left[ \partial _{\mu}\partial_{\rho}h^{\rho}_{\nu}-\partial _{\nu}\partial_{\rho}h^{\rho}_{\mu}-\partial_{\mu}\partial_{\nu}h-\Box h_{\mu \nu}\right] + \mathcal{O}(h^2)\\
    R&=&g^{\mu \nu}R_{\mu \nu}=\kappa(\partial_{\mu}\partial_{\nu}h^{\mu \nu}-\Box h) + \mathcal{O}(h^2),
\end{eqnarray}
where $\Box \equiv \partial^{\mu}\partial_{\mu}$.
Therefore, the action~(\ref{f(R) action in jordan frame}) at first order in $h_{\mu\nu}$ can be written as
\begin{eqnarray}
  S &=& \int \mathrm{d^4 x}\left [\frac{1}{2}h^{\mu \nu}\Box h_{\mu \nu}-h^{\mu \nu}\partial_{\mu}\partial^{\rho}h_{\rho \nu} -\frac{1}{2} h\Box h+\frac{1}{2}h\partial_{\mu}\partial_{\nu}h^{\mu \nu}+\frac{1}{2}h^{\mu \nu}\partial_{\mu}\partial_{\nu}h\right.\nonumber\\
  &&+\left. 2\alpha (h^{\mu \nu} \partial_{\mu} \partial_{\nu} \partial_{\alpha} \partial_{\beta} h^{\alpha \beta} + h \Box^2 h-h^{\mu \nu} \partial_{\mu} \partial_{\nu} \Box h-h \Box \partial_{\mu} \partial_{\nu} h^{\mu \nu}) + \frac{\kappa}{2} h^{\mu \nu} T_{\mu \nu} \right] \label{eq:action}\\
  &=& \int \left[d^4x h_{\mu\nu} \mathcal{E}^{\mu\nu,\alpha \beta} h_{\alpha\beta}+ \frac{\kappa}{2} h^{\mu \nu} T_{\mu \nu} \right], 
\end{eqnarray}
where $T^{\mu\nu}$ is the energy-momentum tensor of the source of the gravitational waves and the kinetic operator $\mathcal{E}^{\mu\nu,\alpha\beta}$ can be written as 
\begin{equation}
    \mathcal{E}^{\mu\nu,\alpha\beta} = \left[\frac{1}{2}P^{(2)\mu\nu,\alpha\beta}-P^{(s)\mu\nu,\alpha\beta}\right]\Box +6 \alpha P^{(s)\mu\nu,\alpha\beta}\Box^2,
    \label{eq:kinetic_operator}
\end{equation}
using the definitions of spin-2 and spin-0 projection operators given by
\begin{eqnarray}
  P^{(2)}_{\mu\nu,\alpha\beta}&=& \frac{1}{2}\left(\theta_{\mu\alpha}\theta_{\nu\beta}+\theta_{\mu\beta}\theta_{\nu\alpha}\right)- \frac{1}{3}\theta_{\mu\nu}\theta_{\alpha\beta} , \\
   P^{(s)}_{\mu\nu,\alpha\beta}&=& \frac{1}{3}\theta_{\mu\nu}\theta_{\alpha\beta},\\
   \theta_{\mu\nu}&=& \eta_{\mu\nu} - \frac{\partial_{\mu}\partial_{\nu}}{\Box}.
\end{eqnarray}
The graviton propagator $D^{(0)}_{\mu\nu,\alpha\beta}$ is the inverse of the kinetic operator $\mathcal{E}^{\mu\nu,\alpha\beta}$ such that
\begin{equation}
    \mathcal{E}^{\mu\nu,\alpha\beta}D^{(0)}_{\mu\nu,\alpha\beta}(x-y)=i\delta^{\mu}_{(\rho}\delta^{\nu}_{\sigma)}\delta^4(x-y).\label{eq:propagator}
\end{equation}

Without appropriate gauge condition it is not possible to inverse the kinetic operator (Eq.~(\ref{eq:kinetic_operator})) in order to get the graviton propagator (Eq.~(\ref{eq:propagator})). To avoid this difficulty, we take an alternative route which is described in the following section. 

\subsection{An equivalent linearized action in Scalar-Tensor form}
Using the conformal transformation between the Jordan frame metric $g_{\mu \nu}$ and the Einstein frame metric $\tilde{g}_{\mu \nu}$: 
\begin{eqnarray}
     g_{\mu \nu}&=&A^{2}(\phi)\tilde{g}_{\mu \nu}, \\
     A^{2}(\phi)&=&\frac{1}{f'(R)}, \\
      f'(R)&=&\frac{df(R)}{dR} ,
\end{eqnarray}
the $f(R)$ gravitational action can be rewritten in scalar-tensor form, in the Einstein frame,
\begin{equation}
    S=\int \sqrt{-\tilde{g}} \mathrm{d^4 x}\left [-\frac{\tilde{R}}{16 \pi G}-\frac{1}{2}\partial_{\mu}\phi \partial^{\mu}\phi + V(\phi)\right ]+S_{M}[ A^{2}(\phi)\tilde{g}_{\mu \nu},\Psi],
    \label{f(R) scalar tensor}
\end{equation}
where the scalar field $\phi$ and the potential $V(\phi)$ are  identified as,
\begin{eqnarray}
     \phi&=&-\sqrt{\frac{3}{16 \pi G}}\ln f'(R), \label{eq: phi def}\\
     V(\phi)&=&\frac{R f'(R)-f(R)}{16 \pi G f'(R)^2}. \label{eq: phipot def}
\end{eqnarray}
Hence the modified equation of motion for the scalar field is,
\begin{eqnarray}
\Box \phi &=& \frac{d V}{d \phi}-\frac{T}{M_{Pl}\beta(\phi)}    
\label{eq:scalar}
\end{eqnarray}
 with $\beta(\phi)=M_{Pl}\frac{d \ln A}{d \phi}$.
 For any $f(R)$ model $\beta(\phi) \equiv 1/\sqrt{6}$. In the non-relativistic and static limit, the Eq.~(\ref{eq:scalar}) becomes
 \begin{equation}
 \begin{split}
     \nabla^2 \phi = & \frac{\text{d}V}{\text{d}\phi} + \frac{\rho}{\sqrt{6}M_{pl}}\\
     = & \frac{\text{d}V_{eff}}{\text{d}\phi}
     \end{split}
 \end{equation}
 where $V_{eff}(\phi;\rho)=V(\phi)+\rho \ln A(\phi)$, $\rho$ being the matter density of the local environment of the scalar field. Therefore, the minima of the effective potential $V_{eff}(\phi)$ depends of the mass density of the local environment and stabilizes around those minima (denoted as $\phi_m(\rho)$) accordingly. Consequently, the mass of the scalar field depends on the matter density of local environment such that $m^2_{\phi}(\rho)=\frac{\text{d}^2V_{eff}}{\text{d}\phi^2}\Big\vert_{\phi_{m}}$. For the dark energy models of $f(R)$ gravity, such as Hu-Sawicki, Tsujikawa, and others, the mass of the scalar field becomes heavier in the high mass density region and lighter in the low mass density region. Thus, the scalar fifth force can be screened in the sufficiently dense environment. This effect is known as the Chameleon Screening \cite{Burrage:2017qrf,Hu:2007nk,PhysRevD.104.084017}.
 
Assuming a static spherically symmetric source object of constant mass density $\rho_0$ and radius $r_s$ embedded in the homogeneous background of matter density $\rho_b$, the scalar field equation can be solved by the method of matching the interior and exterior solutions. The exterior scalar field solution is then given by \cite{smg_gw}
\begin{equation}
    \phi(r)\approx \phi_{\text{VEV}}-M_{Pl}\frac{GM\epsilon}{r}e^{-m_br},  \quad~ r\geq r_s,
    \label{eq:scalar_profile}
\end{equation}
 where $\phi_{VEV}$ is the minima of the scalar field in the background far away from the source object and $m_b$ is the mass of the scalar field at the background matter density $\rho_b$, $M$ is the mass of the source object, and $\epsilon $ is the screened parameter given by
 \begin{equation}
     \epsilon = \frac{\phi_{\text{VEV}}-\phi_0}{M_{Pl}\Phi_N},
 \end{equation}
 where $\phi_0$ is the minima of the scalar field inside the source object and $\Phi_N=GM/r_s$ is the Newtonian surface gravitational potential of the source object. Usually, the matter density of the source object is much higher as compared to the matter density of the background. In our case, we assume compact objects embedded either in the cosmological background or in the galactic background. In such situation, $\phi_{\text{VEV}}>>\phi_0 $ and therefore
 \begin{equation}
     \epsilon =\frac{\phi_{\text{VEV}}}{M_{Pl}\Phi_N}= -\sqrt{\frac{3}{2}}\frac{\ln f'(R_{\text{VEV}})}{\Phi_N}.
 \end{equation}
 From Eq.~(\ref{eq:scalar_profile}) we identify the scalar charge associated to a source object as
 \begin{equation}
     Q= 4\pi M_{Pl} G M \epsilon.
     \label{eq:scalar_charge}
 \end{equation}
The scalar fifth force between two objects with scalar charges $Q_a$ and $Q_b$, separated by the distance $r$ is thus given by 
\begin{equation}
    \begin{split}
        F_5(r)=& - \frac{1}{4\pi} \frac{Q_a Q_b}{r^2}e^{-m_b r}\\
        =& -\frac{GM_aM_b \epsilon_a \epsilon_b}{2r^2} e^{-m_b r}.
    \end{split}
\end{equation}
From the Cassini mission, the solar system bound on the PPN parameter leads to the upper bound on the screened parameter $\vert\epsilon \vert < 2.3\times 10^{-5}$ \cite{Will:2014kxa}. For compact objects like Neutron star and White Dwarf, this upper bound on $\epsilon$ will be more stringent as the surface gravitational potential for compact objects is much higher than the Sun. Therefore, we neglect the effect of fifth force (as $F_5/F_N\approx \mathcal{O}(\epsilon^2)$) in deriving the Keplerian orbit of the binary compact stars.

 We treat the gravitational field $h_{\mu\nu}$ as the small perturbation over the flat background such that
 \begin{equation}
     \tilde{g}_{\mu \nu} = \eta_{\mu \nu}+\kappa h_{\mu \nu}.
 \end{equation}
 Then we linearize the action in Eq.~(\ref{f(R) scalar tensor}) considering one term at a time:
 \begin{eqnarray}
 \sqrt{-\tilde{g}}&=&1+\frac{\kappa}{2}h+\frac{\kappa^2}{8}h^2-\frac{\kappa^2}{4}h^{\alpha \beta}h_{\alpha \beta}+\mathcal{O}(\kappa^3), \\
 \sqrt{-\tilde{g}}\frac{\tilde{R}}{16 \pi G} &\simeq& \frac{1}{2}h^{\alpha \beta} \Box h_{\alpha \beta}-\frac{1}{2}h \Box h + \partial_{\alpha}h^{\alpha \beta} \partial_{\mu}h^{\mu}_{\beta}-\partial_{\alpha}h \partial_{\mu}h^{\alpha \mu}+...,\\
 \sqrt{-\tilde{g}}\partial_{\mu}\phi \partial^{\mu}\phi &\simeq& (1+\frac{\kappa}{2}h+...)(\eta^{\mu \nu}-\kappa h^{\mu \nu})\partial_{\mu}\phi \partial_{\nu}\phi \nonumber \\
 &\simeq& (\partial_{\mu}\phi )^2, \\
 \sqrt{-\tilde{g}}V(\phi)&\simeq& \left(1+\frac{\kappa}{2}h\right)V_{\text{VEV}}+ \frac{\kappa}{2\sqrt{6}}\rho_{\text{VEV}}\left(\phi- \phi_{\text{VEV}}\right) +\frac{1}{2}m^2_{\phi} (\phi -\phi_{\text{VEV}})^2 + ...,\nonumber\\
 \end{eqnarray}
 where
 \begin{eqnarray}
 V_{\text{VEV}}=V(\phi_{\text{VEV}}),\\
 \phi_{\text{VEV}}\simeq -\sqrt{\frac{3}{16 \pi G}}\ln f'(R_{\text{VEV}}).
 \end{eqnarray}
 Note that the effective cosmological constant $\Lambda_{eff}= V_{\text{VEV}}/M_{Pl}^2$, $m^2_{\phi}=\frac{d^2 V}{d \phi^2}\Big|_{\phi_{\text{VEV}}} $, and $\bar{\phi}=\phi-\phi_{\text{VEV}}$. We have assumed flat background, instead of the cosmological FRW background as the time scale of cosmological evolution is much larger than the time scale of evolution of compact binary orbits. Further, we can safely neglect the cosmological density and the cosmological constant as compare to the astrophysical density and, therefore, the scalar field part of the action becomes
\begin{eqnarray}
    S_{\phi}&=& \int \sqrt{-\tilde{g}} \mathrm{d^4 x}\left [\frac{1}{2}\partial_{\mu}\phi \partial^{\mu}\phi- V(\phi) \right ] \nonumber \\
    &\simeq& \int\mathrm{d^4 x}\left [\frac{1}{2}\partial_{\mu}\bar{\phi} \partial^{\mu}\bar{\phi} - \frac{1}{2}m^2_{\phi}\bar{\phi}^2 \right ].
\end{eqnarray}
 Now moving to the matter action $S_{M}[ A^{2}(\phi)\tilde{g}_{\mu \nu},\Psi]$, we have
 \begin{eqnarray}
     S_{M}[ A^{2}(\phi)\tilde{g}_{\mu \nu},\Psi]=\int \sqrt{-g}\mathcal{L}(g_{\mu \nu},\Psi)\mathrm{d^4 x}.
 \end{eqnarray}
 We can expand the Lagrangian in Taylor series,
 \begin{eqnarray}
 \mathcal{L_M}(g_{\mu \nu},\Psi)=\mathcal{L_M}(\bar{g}_{\mu \nu}, \Psi)+\frac{\partial \mathcal{L_M}}{\partial \bar{g}_{\mu \nu}}(g_{\mu \nu} - \bar{g}_{\mu \nu}) +...
 \label{L in Taylor series}
 \end{eqnarray}
 where $\bar{g}_{\mu\nu}$ is the background Jordon frame metric.
 Also expanding $A^2({\phi})$ in Taylor series we can write the Jordan frame metric as,
 \begin{eqnarray}
 g_{\mu \nu} \simeq A^{2}_{\text{VEV}}\eta_{\mu \nu}+\kappa A^2_{\text{VEV}}h_{\mu\nu}+\chi \eta_{\mu \nu}\bar{\phi}+\kappa \chi h_{\mu \nu}\bar{\phi}+...
 \label{g in taylor series}
 \end{eqnarray}
 where $A^2_{\text{VEV}}=A^2(\phi_{\text{VEV}})=1/f'(R_{\text{VEV}})$
 and $\chi =\frac{d}{d\phi}A^2(\phi)\Big|_{\phi_{\text{VEV}}}$. Thus we identify that
 \begin{eqnarray}
 \bar{g}_{\mu \nu}&=&A^2_{\text{VEV}}\eta_{\mu \nu}, \\
 \mbox{and, therefore,} && \nonumber \\
 \sqrt{-g}\mathcal{L_M}(g_{\mu \nu},\Psi)&=& \left(A^4_{\text{VEV}}+\frac{d}{d\phi}A^4(\phi)\Big|_{\phi_{\text{VEV}}}\left(\phi-\phi_{\text{VEV}}\right)+...\right)\left(1+\frac{\kappa}{2}h+...\right) \nonumber \\ && \times \left[ \mathcal{L}_M(\bar{g}_{\mu \nu}, \Psi)+\frac{\partial \mathcal{L_M}}{\partial \bar{g}_{\mu \nu}}\left(\chi \bar{\phi}\eta_{\mu \nu}+\kappa A^2_{\text{VEV}}h_{\mu \nu}+\kappa \chi \bar{\phi}h_{\mu \nu}+...\right) \right]. \nonumber \\
 \end{eqnarray}
 Now using the definition of the energy-momentum tensor,
 \begin{eqnarray}
 T^{\mu \nu}&=& -\frac{2}{\sqrt{-g}}\frac{\partial ( \sqrt{-g}\mathcal{L}_M)}{\partial g_{\mu \nu}}, \nonumber
 \end{eqnarray}
 we write the energy-momentum tensor for background Jordon frame metric,
 \begin{eqnarray}
 \bar{T}^{\mu \nu} = -\left[ 2 \frac{\partial \mathcal{L}_M}{\partial \bar{g}_{\mu \nu}} + A^{-2}_{\text{VEV}}\eta^{\mu \nu}\mathcal{L}_M \right].
 \end{eqnarray}
 Hence, we get,
 \begin{eqnarray}
   \frac{\partial \mathcal{L}_M}{\partial \bar{g}_{\mu \nu}} =-\frac{1}{2}\left(\bar{T}^{\mu \nu}+A^{-2}_{\text{VEV}}\eta^{\mu \nu}\mathcal{L}_M(\bar{g}_{\mu \nu},\Psi) \right).
 \end{eqnarray}
 In the Einstein frame, the background metric is Minkowski, i.e. $\tilde{g}_{\mu\nu}=\eta_{\mu\nu}$ and the corresponding Energy-Momentum tensor is $\tilde{T}_{\mu\nu}$. 
 Therefore,
 \begin{eqnarray}
 \tilde{T}^{\mu \nu}&=&-\left(2 \frac{\partial \mathcal{L}_M}{\partial \eta_{\mu \nu}}+\eta^{\mu \nu}\mathcal{L}_M\right),\\
 \bar{T}^{\mu \nu}&=&A^{-2}_{\text{VEV}}\tilde{T}^{\mu \nu},\\
 \text{and we get,} \nonumber \\
 \frac{\partial \mathcal{L}_M}{\partial \bar{g}_{\mu \nu}}&=&-\frac{1}{2}A^{-2}_{\text{VEV}}(\tilde{T}^{\mu \nu}+\eta^{\mu \nu} \mathcal{L}_M).
 \end{eqnarray}
Thus we get,
\begin{eqnarray}
\sqrt{-g}\mathcal{L}_M (g_{\mu \nu},\Psi)&\simeq& A^4_{\text{VEV}}\bar{L}_M-\frac{1}{2}A^2_{\text{VEV}}\chi \tilde{T}\bar{\phi}-A^4_{\text{VEV}}\frac{\kappa}{2}\tilde{T}^{\mu \nu}h_{\mu \nu}+...
\end{eqnarray}
and hence we obtain the linearized $f(R)$ action,
\begin{eqnarray}
 S_f&=& \int  \mathrm{d^4 x} \left[ -\frac{1}{2}(\partial_{\mu}h_{\nu \rho})^2+\frac{1}{2}(\partial_{\mu}h)^2-(\partial_{\mu}h)(\partial^{\nu}h^{\mu}_{\nu})+(\partial_{\mu}h_{\nu \rho})(\partial^{\nu}h^{\mu \rho}) \right. \nonumber \\ 
 && \left. + \frac{1}{2}A^2_{\text{VEV}}\chi\tilde{T}\bar{\phi}+A^4_{\text{VEV}}\frac{\kappa}{2}\tilde{T}^{\mu \nu}h_{\mu \nu} -\frac{1}{2}\partial_{\mu}\bar{\phi}\partial^{\mu}\bar{\phi}+\frac{1}{2}m^2_{\phi} \bar{\phi}^2 \right].
\end{eqnarray}
Using the definitions of $\phi$ and $\chi(\phi)$, we can rewrite the action,
\begin{eqnarray}
 S_f&=& \int  \mathrm{d^4 x} \left[ -\frac{1}{2}(\partial_{\mu}h_{\nu \rho})^2+\frac{1}{2}(\partial_{\mu}h)^2-(\partial_{\mu}h)(\partial^{\nu}h^{\mu}_{\nu})+(\partial_{\mu}h_{\nu \rho})(\partial^{\nu}h^{\mu \rho}) \right. \nonumber \\ 
 && \left. + \frac{\kappa}{2\sqrt{6}}A^4_{\text{VEV}}\tilde{T}\bar{\phi}+A^4_{\text{VEV}}\frac{\kappa}{2}\tilde{T}^{\mu \nu}h_{\mu \nu} -\frac{1}{2}\partial_{\mu}\bar{\phi}\partial^{\mu}\bar{\phi}+\frac{1}{2}m^2_{\phi} \bar{\phi}^2 \right]
 \label{linearized f(R) action}
\end{eqnarray}

\section{Gravitational radiation from binary system}\label{sec:rad}
In this section we compute the rate of energy loss of the binary system due to gravitational radiation using the linearized action in Eq.~(\ref{linearized f(R) action}). 
The net gravitational emission rate is thus sum of the massless spin-2 graviton radiation and the scalar radiation. Therefore, following the method described in \cite{Mohanty:1994yi,Mohanty:2020pfa,Poddar2022}, the net emission rate is,

\begin{equation}
d \Gamma = d \Gamma^h +d \Gamma^{\phi}
\end{equation}
where the spin-2 graviton emission rate is given by
\begin{eqnarray}
d \Gamma^{h} &=& \frac{\kappa^2}{4}A^{8}_{\text{VEV}} \sum_{\lambda=1}^{2}|T_{\mu \nu}(k')\epsilon_{\lambda}^{\mu \nu}(k)|^2 2 \pi \delta(\omega-\omega')\frac{d^3 k}{(2 \pi)^3}\frac{1}{2 \omega} \nonumber \\
&=& \frac{\kappa^2 A^{8}_{\text{VEV}}}{8 (2 \pi)^2}\sum_{\lambda=1}^{2}(T_{\mu \nu}(k')T^{*}_{\alpha \beta}(k')\epsilon_{\lambda}^{\mu \nu}(k)\epsilon^{*\alpha \beta}_\lambda(k))\frac{d^3 k}{\omega}\delta (\omega-\omega').
\end{eqnarray}
Here $T_{\mu\nu}(k^{\prime})$ is the stress-energy tensor in the momentum space of the spin-2 gravitons. Using the polarization sum of massless spin-2 gravitons,

\begin{eqnarray}
\sum_{\lambda=1}^{2}\epsilon_{\mu \nu}^{\lambda}(k)\epsilon_{\alpha \beta}^{* \lambda}(k)=\frac{1}{2}(\eta_{\mu \alpha}\eta_{\nu \beta}+\eta_{\mu \alpha}\eta_{\nu \alpha})-\frac{1}{2}\eta_{\mu \nu}\eta_{\alpha \beta},
\end{eqnarray}
we get,

\begin{eqnarray}
d \Gamma^{h} = \frac{\kappa^2 A^{8}_{\text{VEV}}}{(2 \pi)^2}\frac{\pi}{5}\left(T_{ij}(\omega')T^{*}_{ji}(\omega')-\frac{1}{3}|T^{i}_{i}(\omega')|^2\right)\delta(\omega-\omega')\omega \delta \omega.
\end{eqnarray}
Therefore , the rate of energy loss due to spin-2 graviton radiation is,
\begin{eqnarray}
\frac{d E^h}{dt}=\frac{\kappa^2 A^{8}_{\text{VEV}}}{20 \pi}\int \left(T_{ij}(\omega')T^*_{ji}(\omega')-\frac{1}{3}|T_{i}^{i}(\omega')|^2\right)\omega^2 \delta(\omega-\omega')d\omega.
\end{eqnarray}

The classical energy-momentum tensor for the binary system orbiting in the $x-y$ plane is~\cite{Mohanty:1994yi,Mohanty:2020pfa, Poddar2022},
\begin{eqnarray}
T_{\mu \nu}(x')&=&\mu \delta^3(\vec{x'}-\vec{x}(t))U_{\mu}U_{\nu}, \\
 \mu &=& \frac{M_1 M_2}{M_1 + M_2} ,\ \text{and} \ U_{\mu} =(1, \dot{x},\dot{y},0),
\end{eqnarray}
where $M_1$ and $M_2$ are the masses of the binary stars, $\mu$ is the reduced mass, $U_{\mu}$ is the non-relativistic four velocity of the reduced mass. 

We can write the Keplerian orbit in the parametric form as,
\begin{equation}
x=a(\cos\xi-e), \hspace{0.4cm} y=a\sqrt{(1-e^2)}\sin\xi, \hspace{0.4cm} \Omega t=\xi-e\sin\xi,
\label{eq:8}
\end{equation}
where $a$ and $e$ are the semi-major axis and eccentricity of the elliptic orbit, respectively. Since the angular velocity of an eccentric orbit is not constant, we can write the Fourier transform of the current density in terms of the $n$ harmonics of the fundamental frequency $\Omega=\Big[G\frac{(M_1+M_2)}{a^3}\Big]^\frac{1}{2}$. Using Eq.~(\ref{eq:8}) we can write the Fourier transforms of the velocity components in the Kepler orbit as,
\begin{equation}
\dot{x}_n=\frac{1}{T}\int ^T_0 e^{in\Omega t}\dot{x}dt=-ia\Omega J^\prime_n(ne),
\label{eq:9}
\end{equation}  
and
\begin{equation}
\dot{y}_n=\frac{1}{T}\int ^T_0 e^{in\Omega t}\dot{y}dt=\frac{a\sqrt{(1-e^2)}}{e}\Omega J_n(ne),
\label{eq:10}
\end{equation}
where we have used $T=2\pi/\Omega$ and the Bessel function identity $J_n(z)=\frac{1}{2\pi}\int^{2\pi}_0 e^{i(n\xi-z\sin\xi)}d\xi$. The prime over the Bessel function denotes the derivative with respect to the argument. Hence the Fourier transforms of the orbital coordinates become,
\begin{equation}
x_n=\frac{\dot{x}_n}{-i\Omega n}=\frac{a}{n}J^\prime_n (ne),\hspace{0.4cm} y_n=\frac{\dot{y}_n}{-i\Omega n}=\frac{ia\sqrt{1-e^2}}{ne}J_n(ne).
\label{eq:11}
\end{equation}
  The Fourier transforms of components of the stress energy tensor with $\omega^\prime=n\Omega$ are given by \cite{Mohanty:2020pfa,Poddar2022},
\begin{eqnarray}
T_{ij}(\omega')T^{ij*}(\omega')-\frac{1}{3}|T^{i}_{i}(\omega')|^2=4 \mu^2 \omega^4 a^4 f(n,e), 
\end{eqnarray}
where,
\begin{equation}
\begin{split}
f(n,e)=\frac{1}{32n^2}\Big\{[J_{n-2}(ne)-2eJ_{n-1}(ne)+2eJ_{n+1}(ne)+\frac{2}{n}J_n(ne)-J_{n+2}(ne)]^2+\\
(1-e^2)[J_{n-2}(ne)-2J_n(ne)+J_{n+2}(ne)]^2+\frac{4}{3n^2}J^2_{n}(ne)\Big\}.
\end{split}
\end{equation}
Therefore,
\begin{eqnarray}
\frac{dE^h}{dt}&=&\frac{32 G}{5}A^{8}_{\text{VEV}}\sum_{n=1}^{\infty}(n \Omega)^2 \mu^2 a^4 (n \Omega)^4 f(n,e)\nonumber\\
&=&\frac{32 G}{5}A^{8}_{\text{VEV}}\mu^2 a^4 \Omega^6 ( 1-e^2)^{-7/2}\left(1+\frac{73}{24}e^2+\frac{37}{96}e^4 \right).
\end{eqnarray}
Thus the emission rate of spin-2 gravitons is exactly same as the Peter-Mathews formula \cite{Peters:1963ux} with the multiplication factor $A^8_{\text{VEV}}$. 
For the scalar part of the gravitational radiation we identify the scalar interaction Lagrangian in the linearized action, Eq.~(\ref{linearized f(R) action}). In the non-relativistic limit the trace of the energy-momentum tensor $\tilde{T}=-\rho$, where $\rho$ is the energy density of the binary systems. Since $\rho$ contains the scalar charges of the source objects, they interact with the scalar field and radiate energy. Therefore, the effective interaction Lagrangian is,
\begin{equation}
\begin{split}
    \mathcal{L}_s= & \frac{\kappa A^4_{\text{VEV}}}{2\sqrt{6}}\bar{\phi}\rho\\
                 \simeq & \frac{\kappa A^4_{\text{VEV}}}{2}\bar{\phi}\rho_s M_{Pl},
\end{split}
\end{equation}
where the scalar charge density $\rho_s$ is related to the energy density $\rho$ as $\rho \equiv \sqrt{6} M_{Pl} \rho_s $. The scalar charge density $\rho_s(x)$ for the binary stars (denoted by $a=1, \, 2$) may be written as,
\begin{equation}
    \rho_s(x)= \sum_{a=1,2} Q_a \delta^3(\vec{x}-\vec{x_a}(t)),
    \label{charge density}
\end{equation}
where $Q_a$ is the scalar charge, Eq.~(\ref{eq:scalar_charge}), in the star and $\vec{x_a}(t)$ represents the Keplerian orbit of the binary stars.
Thus the scalar part of the gravitational radiation is,

\begin{eqnarray}
d \Gamma^{\phi} &=& \frac{\kappa^2A^8_{\text{VEV}}}{4} M_{Pl}^2 \int \vert  \rho_s(\omega')\vert^2 (2\pi)\delta(\omega-\omega')\frac{d^3 k}{(2 \pi)^3 (2\omega)},
\end{eqnarray}
where $d^3 k=k^2 dk d\Omega_{k}$. Using the dispersion relation for the scalar field is $k^2= \omega^2-m^2_{\phi}$, we get,

\begin{equation}
d \Gamma^{\phi}=\frac{\kappa^2 A^{8}_{\text{VEV}}M_{Pl}^2}{32 \pi^2}\int |\rho_s(\omega')|^2\omega\delta(\omega-\omega')\sqrt{1-m_{\phi}^2/{\omega^2}}\, d\omega \, d\Omega_{k},
\label{eq:emission_scalar}
\end{equation}
where $\rho_s(\omega')$ is the Fourier transform of the charge density $\rho_s(s)$ (Eq.~(\ref{charge density})) given by,
\begin{equation}
    \rho_s(\omega')= \int \frac{1}{T} \int^T_0 e^{i\vec{k}\cdot \vec{x}} e^{-i\omega' t} \sum_{a=1,2} Q_a\delta^3(\vec{x}-\vec{x}_a(t)) d^3x dt,
\end{equation}
where $T=2\pi/\Omega$.
Using the Taylor's series expansion $e^{i\vec{k}\cdot\vec{x}}= 1+ i\vec{k}\cdot\vec{x} +...$ and keeping only the leading order contribution, we get,
\begin{equation}
    \rho_s(\omega')= (Q_1+Q_2)\delta(\omega') + i \mu \left(\frac{Q_1}{M_1}-\frac{Q_2}{M_2}\right)\left(k_x x(\omega')+k_yy(\omega')\right) +\mathcal{O}(k^2).
\end{equation}
The first term does not contribute in the radiation formula. Using the Fourier transform $x(\omega')$ and $y(\omega')$ [Eq.~(\ref{eq:11})] and the angular average $<k_x^2>=<k_y^2>=\frac{1}{3}(n\Omega)^2(1-n_0^2/n^2)$, where $n_0=m_{\phi}/\Omega$, we obtain,
\begin{equation}
    |\rho_s(\omega)|^2= \frac{\mu^2}{3}\left(\frac{Q_1}{M_1}-\frac{Q_2}{M_2}\right)^2\Omega^2 a^2 \left(1-\frac{n_0^2}{n^2}\right)\left[J'_n(ne)^2+\frac{1-e^2}{e^2}J_n^2(ne)\right].
    \label{charge density sq FT}
\end{equation}
Using Eq.~(\ref{charge density sq FT}) in Eq.~(\ref{eq:emission_scalar}) we obtain the rate of change of energy loss due to the scalar dipole radiation,
\begin{equation}
\begin{split}
    \frac{dE_s}{dt}= & \frac{\kappa^2 M_{Pl}^2A^8_{\text{VEV}}}{24\pi} \left(\frac{Q_1}{M_1}-\frac{Q_2}{M_2}\right)^2\mu^2\Omega^4 a^2 \sum_{n=1}^{\infty} n^2 \left(1-\frac{n_0^2}{n^2}\right)^{3/2}\left[J'_n(ne)^2+\frac{1-e^2}{e^2}J_n^2(ne)\right]\\
    =& \frac{GA^8_{\text{VEV}}}{3} (\epsilon_1 -\epsilon_2)^2\mu^2 \Omega^4 a^2 \sum_{n=1}^{\infty} n^2 \left(1-\frac{n_0^2}{n^2}\right)^{3/2}\left[J'_n(ne)^2+\frac{1-e^2}{e^2}J_n^2(ne)\right],
    \end{split}
    \label{eq:energy loss dipole}
\end{equation}
 where we used the definition of the scalar charge $Q$ as mentioned in Eq.~(\ref{eq:scalar_charge}). Further, using the definition of the screened parameter $\epsilon$, we rewrite the Eq.~(\ref{eq:energy loss dipole}) as,
 \begin{equation}
 \begin{split}
     \frac{d E_s}{dt}=&\frac{GA^8_{\text{VEV}}}{2} \left(\frac{\ln f'(R_{\text{VEV}})}{\Phi_{N,2}} -\frac{\ln f'(R_{\text{VEV}})}{\Phi_{N,1}}\right)^2\mu^2 \Omega^4 a^2 \\
     & \times \sum_{n=1}^{\infty} n^2 \left(1-\frac{n_0^2}{n^2}\right)^{3/2}\left[J'_n(ne)^2+\frac{1-e^2}{e^2}J_n^2(ne)\right].
     \end{split}
     \label{eq:energy loss diplole gen}
 \end{equation}
 For dark energy models, the mass of the scalar field $m_{\phi}\sim 10^{-33}$ eV at the cosmological scales and for the compact binary systems the typical value of $\Omega \sim 10^{-19}$ eV. Therefore, if $m_{\phi}<<\Omega$, i.e. $n_0<< 1$ and $\left(1-\frac{n_0^2}{n^2}\right)^{3/2}\sim 1$, then using the identity $\sum_{n=1}^{\infty} n^2 \left[J'_n(ne)^2+\frac{1-e^2}{e^2}J_n^2(ne)\right]= (1/4)(2+e^2)(1-e^2)^{-5/2}$ \cite{Peters:1963ux}, we can rewrite the Eq.~(\ref{eq:energy loss diplole gen}) in a more compact form,

\begin{equation}
     \frac{d E_s}{dt}=\frac{GA^8_{\text{VEV}}}{4} \left(\frac{\ln f'(R_{\text{VEV}})}{\Phi_{N,2}} -\frac{\ln f'(R_{\text{VEV}})}{\Phi_{N,1}}\right)^2\mu^2 \Omega^4 a^2 \left[\frac{1+e^2/2}{(1-e^2)^{5/2}}\right].
 \end{equation}
 However, we use Eq.~(\ref{eq:energy loss diplole gen}) to analyse the observation and constraining the theory as the dipole radiation can also occur even when $m_{\phi}<\Omega$ but $m_{\phi}>> 10^{-33}$ eV at the much smaller astrophysical length scale (such as the galactic scale).
The orbital period decay is given by,
\begin{equation}
    \dot{P}_{b}=- 6 \pi G^{-3/2}(M_{1}M_{2})^{-1}(M_{1}+M_{2})^{-1/2}a^{5/2}\dot{E}
    \label{pbdot f}
\end{equation}
where $\dot{E}=\dot{E}_{h}+\dot{E}_{s}$.

\section{Constraints from observations}\label{sec:constr}
 In this section, using the formulae derived in the previous section, we compare the observed period decay of the compact binary systems (neutron star-neutron star/white dwarf) with that predicted theoretically in three $f(R)$ gravity models, namely, the Hu-Sawicki model~\cite{Hu:2007nk}, the Starobinsky~\cite{Starobinsky:2007hu} model and the Tsujikawa model~\cite{PhysRevD.77.023507}. We constrain the parameters of these three models and compare these constraints with the solar system constrains~\cite{Will:2005va}. We use four binary objects for our analysis, i.e., PSR B1913+16 (Hulse-Taylor binary)~\cite{Weisberg:2016jye}, PSR J1141-6545 (High-eccentric NS-WD binary)~\cite{Tauris:1999wm,PhysRevD.66.024040,Bhat:2008ck,Verbiest:2012ie}, PSR J1738+0333 (Low-eccentric NS-WD binary)~\cite{Freire:2012mg}, and PSR J0348+0432 (Low-eccentric NS-WD binary)~\cite{Antoniadis:2013pzd} listed in Table~\ref{tab:objects}.
 
We define relative `change' of intrinsic (i.e. observed) orbital period decay and $f(R)$ predicted orbital period decay with respect to that predicted from GR as,
\begin{eqnarray}
\Delta_{\text{Obs}}&=&\left|\frac{\dot{P}_{b,\text{Intrinsic}}-\dot{P}_{b,\text{GR}}}{\dot{P}_{b,\text{GR}}}\right| ,\\
\Delta_{\text{f(R)}}&=&\left|\frac{\dot{P}_{b,\text{f(R)}}-\dot{P}_{b,\text{GR}}}{\dot{P}_{b,\text{GR}}}\right|.
\label{Delta f}
\end{eqnarray}
The allowed $f(R)$ models must satisfy the condition $\frac{\Delta_{f(R)}}{\Delta_{Obs}}<1$. 

\begin{table}[h]
\centering
\resizebox{\linewidth}{!}{%
\begin{tabular}{ |l|c|c|c|c|c| }
 
 \hline
Parameters \hspace{0.01cm} & \textbf{PSR B1913+16}\hspace{0.01cm}&\textbf{PSR J1738+0333}\hspace{0.01cm}&\textbf{PSR J1141-6545}\hspace{0.01cm}&\textbf{PSR J0348+0432}\hspace{0.01cm}\\
 \hline
\textbf{Pulsar mass} $m_1$ ($M_{\odot}$) &$1.438\pm 0.001$ &$1.46^{+0.06}_{-0.05}$ &$1.27\pm0.01$ &$2.01 \pm 0.04$ \\
\textbf{Companion mass} $m_2$ ($M_{\odot}$)&$1.390\pm 0.001$ & $0.181^{+0.008}_{-0.007}$ &$1.02\pm0.01$ &$0.172 \pm 0.003$\\
\textbf{Eccentricity} $e$ &$0.6171340(4)$ &$(3.4\pm 1.1)\times10^{-7}$ &$0.171884(2)$ &$10^{-6}$\\
\textbf{Orbital period} $P_b$ (d)&$0.322997448918(3)$&$0.3547907398724(13)$ &$0.1976509593(1)$ &$0.102424062722(7)$\\
\textbf{Intrinsic} $\dot{P_b}(10^{-12}\rm{ ss^{-1}})$ &$-2.398\pm 0.004$ &$(-25.9\pm 3.2)\times 10^{-3}$ &$-0.403(25)$ &$(-27.3 \pm 4.5) \times 10^{-2}$\\
\textbf{GR} $\dot{P_b}(10^{-12}\rm{ ss^{-1}})$ &$-2.40263\pm 0.00005$&$-27.7^{+1.5}_{-1.9}\times 10^{-3}$ &$-0.386$ &$-25.9 \times 10^{-2}$\\
 \hline
\end{tabular}
}
\caption{Summary of the measured orbital parameters and the orbital period derivative values from observation and GR for PSR B1913+16~\cite{Weisberg:2016jye}, PSR J1738+0333~\cite{Freire:2012mg}, PSR J1141-6545~\cite{Bhat:2008ck}, and PSR J0348+0432~\cite{Antoniadis:2013pzd}. The uncertainties in the last digits are quoted in the parenthesis.}
\label{tab:objects}
\end{table}

\subsection{Hu-Sawicki Model}
The Hu-Sawicki dark-energy model~\cite{Hu:2007nk} is given by
\begin{equation}
f(R)= R - m^2\frac{c_1 \left(R/ m^2\right)^{n}}{c_2 \left(R/ m^2\right)^{n} +1},
\label{eq:hu_sawicki}
\end{equation}
where 
$n>0$, $c_1>0$, and $c_2>0$ are dimensionless parameters and $m^2$ is the mass scale in the theory defined by
$m^2= 8\pi G \bar{\rho}_0/3$, $\bar{\rho}_0$ being the present day average matter density of the Universe. The theory is designed such that it effectively approaches the $\Lambda CDM$ model in the high curvature/ redshift limit and can also explain the present day observed accelerated expansion of the Universe, i.e. $\lim_{R \to \infty} f(R)$ $= R - 2\Lambda_{eff}$  and $\lim_{R \to 0} f(R) =0$ . The second condition is satisfied if $n>0$. For large $R$, the model, Eq.~(\ref{eq:hu_sawicki}), can be approximated as,
\begin{equation}
    f(R) = R - m^2\frac{c_1}{c_2} + m^2\frac{c_1}{c_2^2}\left(\frac{m^2}{R}\right)^n + \mathcal{O}(R^{-2n}).
    \label{eq: hu sawicki approx}
\end{equation}
 Then the effective cosmological constant is identified as $\Lambda_{eff}= \frac{m^2 c_1}{2c_2}$ and
 \begin{equation}
     \frac{c_1}{c_2}= 6 \frac{1-\Omega_{m0}}{\Omega_{m0}},
     \label{eq: c1 by c2}
 \end{equation}
where $\Omega_{m0}=\bar{\rho}_{0}/\rho_{c0}$ is the ratio of the present day matter density and the present day total energy density of the Universe.

\begin{figure}[!htbp]
    \centering
    \subfigure[]{\includegraphics[scale=0.224]{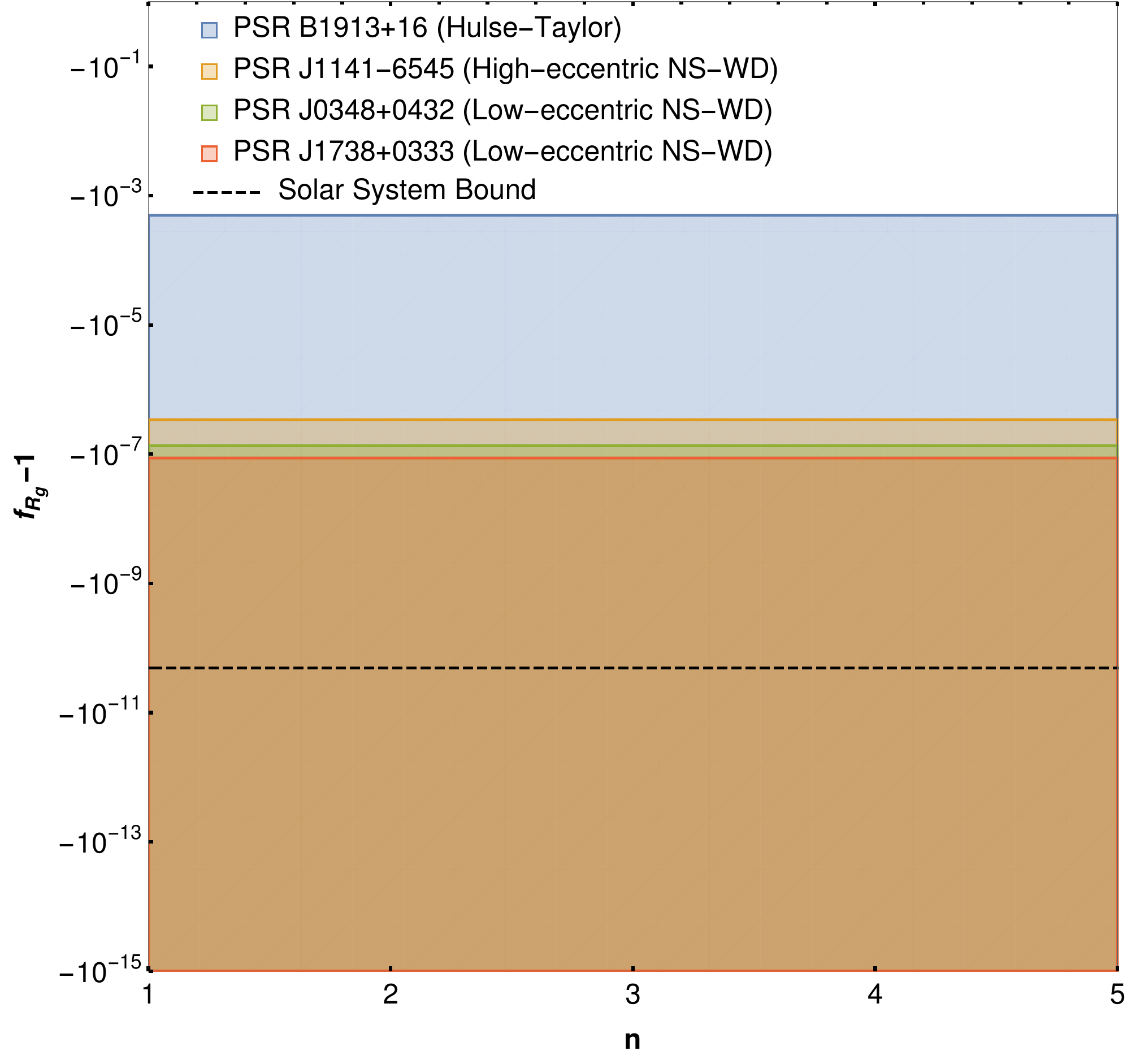}\label{subfig1}}
    \subfigure[]{\includegraphics[scale=0.232]{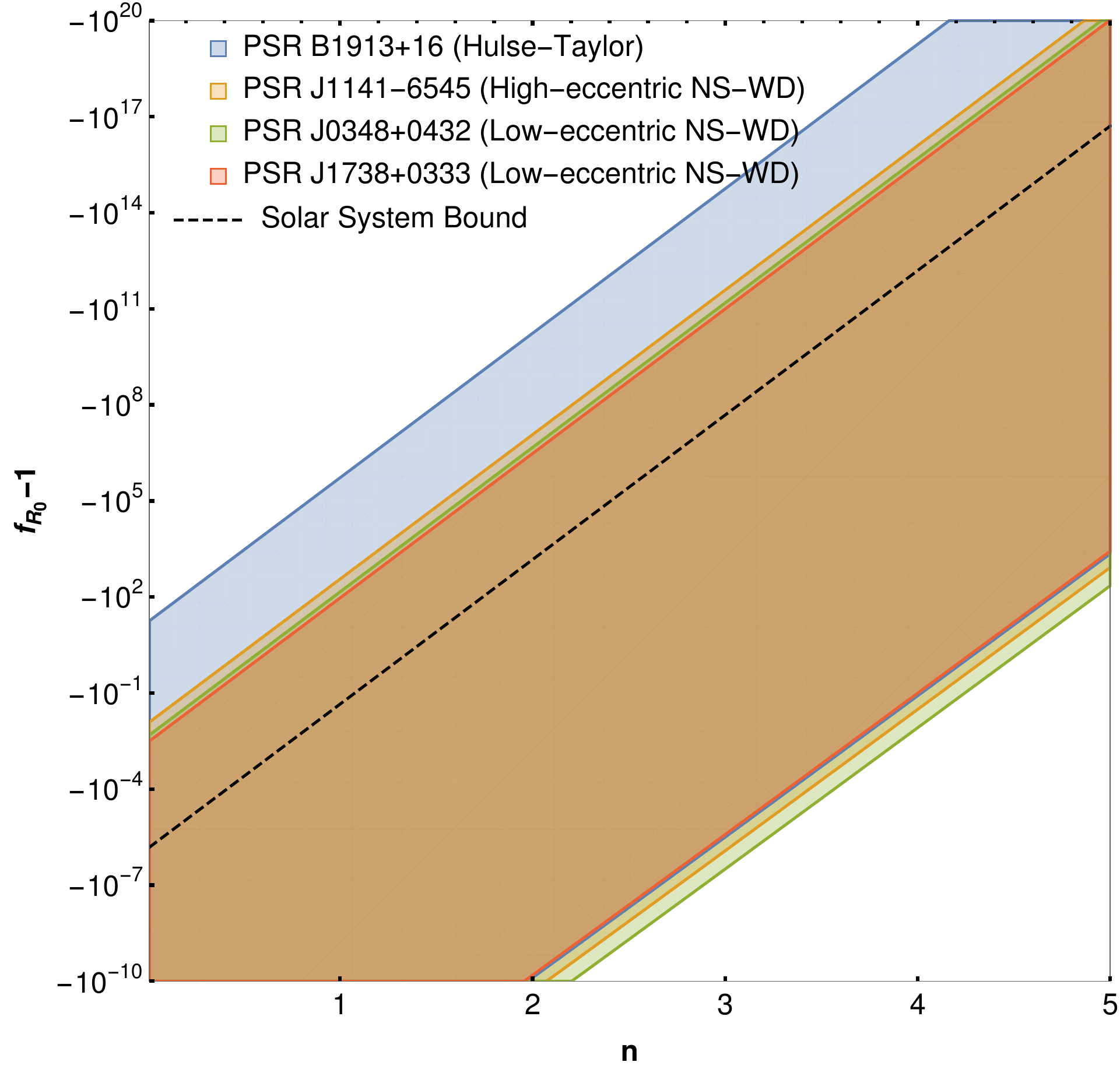}\label{subfig2}}
    \caption{\textit{($a$): }Constraints on galactic scale amplitude $f^{\prime}(R_g)$ is presented in the parameter space $f^{\prime}(R_g)-1$ (denoted as $f_{R_g}-1$) vs. $n$ for the Hu- Sawicki model. The shaded regions are allowed by the corresponding binary stars system. The black dashed line indicates the boundary below which any parameter values are allowed by the Solar-System \cite{Hu:2007nk}. \textit{($b$): }Constrains on cosmological scale amplitude $f^{\prime}(R_0)$ are shown in the $(f^{\prime}(R_0)-1)$ vs. $n$ plane. In the figure, $f^{\prime}(R_0)$ is labeled as $f_{R_0}$.}
    \label{fig:HS_galactic_bounds}
\end{figure}

With the help of the constraint Eq.~(\ref{eq: c1 by c2}), we have now two free parameters in the theory. We choose  $f'(R_{\text{VEV}})$ (the derivative of $f(R)$ with respect to $R$ at the background of the binary system) and $n$.
Assuming the local background of the binary system is the galactic medium where the effective scalar potential $V_{eff}(\phi)$  has a minimum, we have $R_{\text{VEV}}\equiv R_{g}=8\pi G \rho_g$, where $\rho_g= 10^{-24}$ $gm$ $cm^{-3}$ is the mean galactic density. The mass of the scalar field at the galactic background is given by
\begin{equation}
    m^2_{\phi}(\rho_g)=\frac{1}{3}\left[\frac{1}{f''(R_g)}-\frac{4f(R_g)}{f'(R_g)^2}+ \frac{R_g}{f'(R_g)}\right],
    \label{eq: mass gal hu}
\end{equation}
where
\begin{eqnarray}
    f(R_g)&=& R_g\left[ 1- 2\left(\frac{1}{\Omega_{m0}}-1\right)\frac{\bar{\rho}_0}{\rho_g}- \frac{f'(R_g)-1}{n}\right],
    \label{eq: f gal hu}\\
    f''(R_g) &=& - \frac{n+1}{R_g}\left[f'(R_g)-1\right].
    \label{eq: f2prime gal hu}
\end{eqnarray}
Eqs.~(\ref{eq: mass gal hu}), (\ref{eq: f gal hu}), and (\ref{eq: f2prime gal hu}) are obtained from Eqs.(\ref{eq: phi def}), (\ref{eq: phipot def}), and (\ref{eq: hu sawicki approx}). Thus the formulae for rate of energy loss due to scalar dipole radiation, Eq.~(\ref{eq:energy loss diplole gen}), the orbital period decay, Eq.~(\ref{pbdot f}), and the relative change in period decay, Eq.~(\ref{Delta f}) can be written as the function of the free parameters $f'(R_g)$ and $n$. We constrain the parameter space ($f'(R_g), n$) from the observations by imposing the conditions $0<m^2_{\phi}(\rho_g)< \Omega^2$ and $\Delta_{f(R)}< \Delta_{\text{Obs}} $. In Fig.~\ref{subfig1}, we show the constraint on the parameter space. We see that constraint on the Hulse-Taylor binary pulsar ($ \vert f'(R_g) -1  \vert < 10^{-3}$) is the weakest among  all four systems we considered, where as for other two NS-WD systems the constraints are stronger.  For the high eccentric NS-WD binary PSR J1141-6545, $ \vert f'(R_g) -1  \vert < 10^{-6}$ and  for the low eccentric NS-WD binary PSR J1738+0333, $ \vert f'(R_g) -1  \vert < 10^{-7}$. However all the constraints are weaker than the Solar-System bound $ \vert f'(R_g) -1  \vert < 10^{-10}$ which comes from the Cassini Mission \cite{Hu:2007nk}. We note that all the bounds are very weakly dependent on $n$. The reason is that $n$ appears in the dipole radiation formula only through the definition of $m_{\phi}$, whereas the dominating factor is the difference in the charge to mass ratio which does not depend on $n$.   

We map the constraint on the parameter space ($f'(R_g), n$) at the galactic scale to the parameter space ($(f'(R_0)), n $), where $R_0$ is the curvature of the cosmological background at the present epoch. The mapping is done using the relation
\begin{equation}
    \frac{f'(R_g)-1}{f'(R_0)-1}= \left(\frac{R_g}{R_0}\right)^{n+1}= \left[\frac{\rho_g}{\bar{\rho}_0(4/\Omega_{m0}-3)}\right]^{n+1}.
\end{equation}
In Fig.~\ref{subfig2}, we show the constraint on the parameters translated onto the cosmological scale. The shaded regions are allowed ones for each of the binary systems. The lower boundary corresponds to the fact that $m_{\phi}= \Omega$. The Solar-System constraint is bounded below the black dashed line. Here, we note $n$ dependence in the figure.

\subsection{Starobinsky dark energy model}
In the same year of publication of Hu-Sawicki model, Starobinsky also proposed a $f(R)$ dark energy model \cite{Starobinsky:2007hu},
\begin{equation}
    f(R)= R + \lambda R_c\left[\left(1+ \frac{R^2}{R_c^2}\right)^{-k} -1\right],
\end{equation}
where $k, \lambda>0$ and $R_c$  is the curvature scale of the order of the  cosmological constant.
For large $R$, the model can be approximated as,
\begin{equation}
    f(R)= R- \lambda R_c + \lambda R_c \left(\frac{R_c}{R}\right)^{2k} + \mathcal{O}(R^{-2k-2}).
    \label{eq: starobinsky approx}
\end{equation}
Therefore, we identify the effective cosmological constant in the model as $ \Lambda_{eff}= \lambda R_c/2 $. The approximated forms of the Hu-Sawicki model, Eq.~(\ref{eq: hu sawicki approx}), and the Starobinsky model, Eq.~(\ref{eq: starobinsky approx}), are the same with the one to one mapping:
\begin{equation*}
    2k \to n , \quad~   R_c \to C_2^{-1/n}, \quad~ \lambda \to m^2 C_1 C_2^{1/n-1}. 
\end{equation*}
However, this mapping does not hold for full theory and is only relevant at high curvature regime. For our purpose it is sufficient to assume the large $R$ approximated forms and therefore we get exactly same constraint on the Starobinsky model as we get in the Hu-Sawicki model. The only difference is that we need to keep the mapping in the mind.

\subsection{Tsujikawa model}
In the next year of publications of the Hu-Sawicki and the Starobinsky model, another well-known $f(R)$ dark energy model was proposed by Tsujikawa \cite{PhysRevD.77.023507}. The model is given by,
\begin{equation}
    f(R)= R - \lambda R_c \tanh{\frac{R}{R_c}},
\end{equation}
where $\lambda>0$ and $R_c>0$ are only two parameters in the theory. In the limit $R\to \infty$, this theory approaches $\lim_{R\to \infty} f(R)= R- \lambda R_c$. Therefore, here also the effective cosmological constant becomes $\Lambda_{eff}=\lambda R_c/2$. With this constraint relation, the theory has only one free parameter. We choose $\lambda$ to be the free parameter and $R_c = 2\Lambda_{eff}/\lambda$. 

The mass of the scalar field at the galactic scale is given by Eq.~(\ref{eq: mass gal hu}), where $f(R_g)$, $f'(R_g)$, and $f''(R_g)$ are now given by,
\begin{eqnarray}
    f(R_g)&=& R_g\left[1- \frac{2\bar{\rho}_0}{\rho_g}\left(\frac{1}{\Omega_{m0}}-1\right)\tanh{\left[\frac{\lambda \rho_g}{2\bar{\rho}_0(1/\Omega_{m0}-1)}\right]}\right],\\
    f'(R_g)&=& 1- \lambda \,\text{sech}^2\left[\frac{\lambda \rho_g}{2\bar{\rho}_0(1/\Omega_{m0}-1)}\right], \label{eq: fprime gal tsuji}\\
    f''(R_g) &=& -\frac{\lambda \rho_g}{R_g \bar{\rho}_0(1/\Omega_{m0}-1)}\tanh{\left[\frac{\lambda \rho_g}{2\bar{\rho}_0(1/\Omega_{m0}-1)}\right]}\left[ f'(R_g) -1 \right].
\end{eqnarray}

By imposing the conditions $0<m^2_{\phi}(\rho_g)< \Omega^2$ and $\Delta_{f(R)}< \Delta_{\text{Obs}} $ we get the constraint on the parameter $\lambda$ which can be translated into the constraint on  $f'(R_g)-1$ , using the Eq.~(\ref{eq: fprime gal tsuji}). Further the constraint can be translated to the cosmological scale by using the relation,
\begin{equation}
    f'(R_0)= 1-\lambda \, \text{sech}^2\left[\frac{\lambda (4/\Omega_{m0}-3)}{2(1/\Omega_{m0}-1)}\right].
\end{equation}
In the Table~\ref{table2}, we show the constraints on the parameters for three binary systems. We note that for the binary systems  the constraints are of the same order of magnitude. We get $10^{-16}\lesssim \vert f'(R_g)-1\vert \lesssim 10^{-6}$ at the galactic scale and $10^{-11}\lesssim \vert f'(R_0)-1\vert \lesssim 10^{-4}$ at the cosmological scale. 
\begin{table}[h!]
\centering
\resizebox{\linewidth}{!}{%
\begin{tabular}{|c|c|c|c|}
\hline 
System & $\lambda$ & $\vert f_{R_{g}}-1 \vert$ &$\vert f_{R_{0}}-1\vert$ \\ 
\hline 
\textbf{Solar-System} & $\lambda < 4.9 \times 10^{-11}$ or $\lambda > 1.11 \times 10^{-4} $ & $ < 4.9\times 10^{-11}$  & $ < 0.2$ \\ 
\textbf{Hulse-Taylor Binary} & [$1.02 \times 10^{-11}$, $2.09 \times 10^{-4}$] & $[6.74 \times 10^{-16}, 6.21 \times 10^{-6}]$ & $[1.02  \times 10^{-11}, 2.09 \times 10^{-4}]$ \\ 
\textbf{PSR J1141-6545} & [$7.41 \times 10^{-12}$, $2.09 \times 10^{-4}$] & $[6.74 \times 10^{-16}, 3.39 \times 10^{-6}]$ & $[7.41  \times 10^{-12}, 2.09 \times 10^{-4}]$ \\ 
\textbf{PSR J1738+0333} & [$1.10\times10^{-11}$, $2.09 \times 10^{-4}$] & $[6.74 \times 10^{-16}, 8.71 \times 10^{-6}]$ & $[1.10\times 10^{-11}, 2.09 \times 10^{-4}]$ \\ 
\textbf{PSR J0348+0432} & [$4.79\times10^{-12}$, $2.09 \times 10^{-4}$] & $[6.74 \times 10^{-16}, 1.35 \times 10^{-7}]$ & $[4.79\times 10^{-12}, 2.09 \times 10^{-4}]$ \\ 
\hline 
\end{tabular}
}
\caption{Bounds on the Tsujikawa Model parameter for the five binary systems considered. The bound on the Solar-System comes from the Cassini Mission \cite{Hu:2007nk}. }
\label{table2}
\end{table}

\section{Conclusions}\label{sec:conc}
   In this paper we derive the most general formula for the rate of energy loss of quasi-stable compact binary systems in $f(R)$ theories of gravity using a single vertex graviton emission process from a classical source. Then we use it 
to  put constraints on three well known $f(R)$ dark energy models, namely the Hu-Sawicki, the Starobinsky, and the Tsujikawa models from binary pulsar observations. 

In $f(R)$ gravity, an extra massive scalar mode appears apart from the massless spin-2 modes. This extra scalar mode affects the orbital motion of the binary stars in two ways. One is that an attractive
short ranged ``fifth" force adds up to the usual Newtonian gravitational force between two compact objects. The other effect is that the scalar dipole radiation carries away some part of the total mechanical energy of the binary system. The scalar fifth force depends on the product of the scalar charges of the binary stars and this eventually leads to the ratio of the scalar fifth force to the Newtonian force being proportional to the product of the screened parameters ($\epsilon_a \epsilon_b $) for the binary stars. The screened parameter is a $f(R)$-model dependent parameter and bounded by $\vert \epsilon \vert < 10^{-5}$ for the Solar-System. The bound becomes smaller and smaller for the compact objects like white-dwarf and neutron stars as the surface gravitational potential for these objects are much higher than that of the Sun. Therefore, we can safely assume that the quasi-stable orbits of binary compact stars are unaffected by the scalar fifth force. On the other hand, the scalar dipole radiation depends on the square of the difference of the scalar charge to mass ratio of the binary stars, which means effectively the  square of the difference in the screened parameters (i.e. $(\epsilon_a- \epsilon_b)^2$ ). For the binary neutron star systems this effect could be negligible although non-vanishing if there is a slight difference in masses of the companion stars. However, for the neutron star- white dwarf or neutron star - black hole systems there could be significant effect of the scalar dipole radiation. Further it is to be noted that contribution of the scalar dipole radiation rapidly falls off when the scalar mass is greater than the orbital period of the binary system ( i.e. $m_{\phi} > \Omega$). Again the scalar mass depends on the $f(R)$ model and also the background matter density where the effective scalar potential attains a minimum.

Considering all these, we have derived the rate of energy loss due to scalar dipole radiation from the linearized $f(R)$ gravity when written in the scalar-tensor format in the Einstein frame. We have obtained the most general formula, i.e. Eq.~(\ref{eq:energy loss diplole gen}), which applies to arbitrary eccentricity of the binary orbit and arbitrary scalar mass. This formula is different from earlier results in the literature \cite{PhysRevD.54.1474,LIU2018286}, which are applicable under specific limits. Moreover our method is semi-classical while  theirs are fully classical. Earlier, the scalar dipole radiation from binary pulsar systems in Brans-Dicke theory was studied in \cite{PhysRevD.66.024040}. However,  the scalar field is massless in their study. Recently, radiation from eccentric binaries in the classical framework has been studied in Horndeski theories with the massless scalar field and without the screening assumption \cite{PhysRevD.106.064046}.    

Using the general formula for the rate of energy loss, we compare the prediction for the period decay of the binary orbits with that of the Peter-Mathews formula using GR. We have applied our theoretical result to constrain $f(R)$ dark energy models from the observations of the period decay of three binary systems of varied eccentricity and types, which are PSR B1913+16 (Hulse-Taylor), PSR J1141-6545 (high-eccentric NS-WD), PSR J1738+0333 (low-eccentric NS-WD), and PSR J0348+0432 (low-eccentric NS-WD). To summarize our results, at the galactic mean density as the background of the binary system, we found that for Hu-Sawicki model the NS-WD systems give stronger constraints than the Hulse-Taylor binary pulsar which is a NS-NS system. However all these constraints are much weaker than the Solar-System constraint. Although at the Galactic scale the constraints on $f'(R_g)-1$ weakly depend on the model parameter $n$, at the cosmological scale the translated constraint on $f'(R_0)-1$ strongly depends on $n$. The Starobinsky model has the one to one mapping  with the Hu-Sawicki model at the high curvature regime which include our cases in this paper. The Tsujikawa model provides better constraints both at the galactic scale and when translated onto the cosmological scale. One important advantage of the Tsujikawa model over other two is that effectively it has only one free parameter when the theory goes to the $\Lambda CDM$ limit in the high curvature regime and therefore, this model is relatively more robust. Also, from comparison with the observations, it has been noted that all four binary systems provide similar constraints for the Tsujikawa model. The constraints are $\vert f'(R_g)-1\vert< 10 ^{-6}$ at the galactic scale and  $\vert f'(R_0)-1\vert< 10 ^{-4}$ when translated at the cosmological scale. In Table~\ref{table3}, we compare our result with other constraints from different observations at different scales. We note that our bound is stronger than those coming from most of the astrophysical observations and even some cosmological observations, like the CMB spectrum.   

\begin{table}[h]
\centering
\begin{tabular}{lcp{0.23\textwidth}}
\hline
Observations & $\vert f'(R_0)-1\vert$ constraints & Ref. \\
\hline
Solar-System bounds (Cassini mission) & $ \lesssim 0.2$ ${}^{*}$  & \cite{Will:2014kxa} \\
Supernova monopole radiation & $< 10^{-2} $ & \cite{Upadhye:2013nfa} \\
Cluster density profiles (Max-BCG) & $< 3.5\times 10^{-3}$ & \cite{PhysRevD.85.102001} \\
CMB spectrum & $< 10^{-3}$ & \cite{PhysRevD.76.063517}\\
{GW170817 (GW from BNS merger)} &  $< 3\times 10^{-3}$ & \cite{PhysRevD.99.044056} \\
{\bf Period decay of binary system} &  $ \mathbf{< 2.09 \times 10^{-4}}$ ${}^{\dagger}$& {\bf This work}\\
 &   & {\bf (Tsujikawa model)} \\
Cluster abundances & $< 1.6\times 10^{-5}$ & \cite{PhysRevD.80.083505,PhysRevD.92.044009} \\
CMB + BAO + $\sigma_8-\Omega_m$ relationship ${}^{**}$ & $< 3.7\times 10^{-6}$ & \cite{PhysRevD.90.103512}\\
Strong gravitational lensing (SLACS) & $< 2.5\times 10^{-6}$  & \cite{Smith:2009fn} \\
Redshift-space distortions & $< 2.6 \times 10^{-6}$ & \cite{PhysRevD.91.063008} \\
Distance indicators in dwarf galaxies &$ < 5\times 10^{-7}$ & \cite{Jain_2013} \\
\hline
\end{tabular}
\caption{\label{table3} Comparison of the bounds on the cosmological scale amplitude $f'(R_0)$ from different observations.\\ ${}^{*}$ This is obtained for the Tsujikawa model, when translated from the bound $\vert f'(R_{gal})-1\vert \lesssim 5\times 10^{-11}$ at the galactic scale. \\
${}^{\dagger}$ The Tsujikawa model gives the best constraint over other two models: the Hu-Sawicki and the Starobinsky model.\\
${}^{**}$ Taking into account cluster number counts (PSZ catalog) and weak-lensing tomography measurements (CFHTLens). This analysis assumes the Hu-Sawicki model.}
\end{table}

Finally, the systems for observation that we considered here are quasi-stable binary star systems. Observations of gravitational waves from binary merger events by LIGO-Virgo provide the opportunity to test the gravity in the very strong field regime and therefore can give more stringent bound on the $f(R)$ gravity. In our earlier work \cite{PhysRevD.99.044056} we provided one such bound on $f(R)$ models from the inspiral stage (dynamic but far away from the merger stage) of the binary neutron star merger event GW170817. In the future observations, we hope to see some detection of neutron star - black hole or neutron star- white dwarf merger events by LIGO- Virgo which will provide stronger bounds. Also it would be interesting to see if our semi-classical approach can also be applied for GW emission from inspiral or even in the merger stage.  

\section*{Acknowledgements}
Research of SJ is partially supported by the SERB, DST, Govt. of India, through a TARE fellowship
grant no. TAR/2021/000354, hosted by the department of Physics, Indian Institute of Technology Kharagpur.
\bibliographystyle{JHEP}
\bibliography{FRgrav.bib}

\end{document}